\documentclass[aps,twocolumn]{revtex4}
\usepackage{amsmath}
\usepackage{amssymb}
\usepackage{amsthm}
\usepackage{graphics}
\usepackage{eufrak,stmaryrd,upgreek}
\DeclareMathAlphabet{\mathpzc}{OT1}{pzc}{m}{it}

\newcommand\half{\frac{1}{2}}
\newcommand\beq{\begin{equation}}
\newcommand\eeq{\end{equation}}
\newcommand\beqa{\begin{eqnarray}}
\newcommand\eeqa{\end{eqnarray}}

\newcommand\Ao{\hat A}

\newcommand\phio{\hat \phi}
\newcommand\vphio{\hat \varphi}

\newcommand\Xo{\hat X}
\newcommand\Yo{\hat Y}
\newcommand\Zo{\hat Z}

\newcommand\Tc{\mathcal{T}}
\newcommand\Nc{\mathcal{N}}
\newcommand\Wc{\mathcal{W}}
\newcommand\Lc{\mathcal{L}}

\newcommand\Oc{\mathcal{O}}
\newcommand\Cc{\mathcal{C}}
\newcommand\Dc{\mathcal{D}}
\newcommand\hC{\hat{\mathcal{C}}}
\newcommand\hcomp{\hat\complement}
\newcommand\hud{\hat\updelta}
\newcommand\hp{\hat\partial}
\newcommand\bdot{\boldsymbol{\cdot}}

\begin{document}

\title{General Ordering Theorem}   
\author{L. Ferialdi}
\email{luca.ferialdi@gmail.com}
\affiliation{Dipartimento di Fisica e Chimica, Universit$\grave{a}$  degli Studi di Palermo, via Archirafi 36, I-90123 Palermo, Italy}

\begin{abstract}
The problem of ordering operators has afflicted quantum mechanics since its foundation. %is a relevant problem of the mathematical foundations of quantum mechanics. 
Several orderings have been devised, but a systematic procedure to move from one ordering to another is still missing. The importance of establishing relations among different orderings is demonstrated by Wick's theorem (which relates time ordering to normal ordering), which played a crucial role in the development of quantum field theory.
We prove the General Ordering Theorem (GOT), which establishes a relation among any pair of orderings, that act on operators satisfying generic (i.e. operatorial) commutation relations. We expose the working principles of the GOT by simple examples, and we demonstrate its potential by recovering two famous algebraic theorems as special instances: the Magnus expansion and the Baker-Campbell-Hausdorff formula. Remarkably, the GOT establishes a formal relation between these two theorems, and it provides compact expressions for them, unlike the notoriously complicated ones currently known.  %and  them.
\end{abstract}

\maketitle

\section{Introduction}
The Quantum Theory is founded on the mathematical machinery of the Hilbert spaces, which come along with states, operators, etc.
%The Hilbert space structure of the Quantum theory brings along the necessity of working with operators. 
Since operators are non-commuting objects, one needs to be careful to their order  when a product  acts on a state. This fact %is very well known since the early stages of the developments of quantum mechanics, and 
lead to the conception of different ordering (super)operators (henceforth simply  named``orderings'') like Dyson's time (or path) ordering~\cite{Dys49}, normal ordering~\cite{Wic50}, Weyl ordering~\cite{Wey27,Wig32}, and Cahill-Glauber s-orderings~\cite{CahGla69} to name a few. Orderings are rooted so deeply in the quantum theory, that they are involved in several fields of research like e.g. quantum field theory~\cite{Dys49,Wic50,HouKin49,And54,Mat55,Gau60,Vagetal90,EvaSte96,Daletal82}, statistical physics~\cite{Kel65,Hal75}, quantum optics~\cite{CahGla69}, phase space representation~\cite{Wig32,AgaWol69,AgaWol70,AgaWol70b,AgaWol70c}, spin systems~\cite{PerCap77,Peretal84}, quantum chemistry~\cite{KutMuk97,Konetal10}, and Yang-Mills theories~\cite{Sch14}. Nonetheless, to date only few theorems are known that establish connections among different orderings. The most famous is undoubtedly Wick's theorem~\cite{Wic50,Choetal85}, which had a tremendous impact in the development of quantum field theory. Wick's theorem indeed, besides being a theorem that allows to express higher order moments of normal distributions in terms of their covariance matrix, can be understood as a theorem that relates time ordering to normal ordering of bosonic and fermionic operators.  Similar theorems, involving different orderings, were developed in diverse contexts~\cite{Gau60,Vagetal90,EvaSte96,AgaWol69,AgaWol70,AgaWol70b,AgaWol70c,PerCap77,Peretal84,KutMuk97,Konetal10,SilPie82}, but only recently Wick's theorem was generalized to any pair of orderings~\cite{FerDio21}. A common feature shared by these theorems is the they hold only for operators that (anti-)commute to a complex number. A theorem relating general orderings of operators satisfying arbitrary (i.e. operatorial) commutation relations was missing, before the publication of this paper containing the General Ordering Theorem.

Before moving on, we remark that the importance of the General Ordering Theorem goes far beyond the scopes of quantum physics, as ordering issues arise in any mathematical problem involving non-commutative algebras, like e.g. group theory~\cite{Mag50,Bor91,SchLyn01}, analysis of partial differential equations~\cite{FolSte82,Hor67,Nagetal85}, Lie algebras~\cite{Bou89,Ser06,Sep07}, etc. In these contexts two important theorems are known: the Magnus Expansion (ME)~\cite{Mag54,Blaetal}, which allows to express the solution of an operatorial linear differential equation as a pure exponential; and the Baker-Campbel-Hausdorff formula (BCH)~\cite{Cam97,Bak05,Hau06,Dyn47,AchBon12}, which allows to rewrite the product of two operatorial exponentials as a single exponential. Although the ME and the BCH are not usually referred to as ordering theorems, we will next show that they can be understood in this way, thus representing the only two examples (to the author's knowledge) of theorems among orderings of generic (i.e. not c-number-commuting) operators. As such, they also share the merit  of giving a clear understanding of the level of complexity that one has to face when dealing with operators having generic commutation relations, instead of c-number-commuting operators. For example, the BCH can be expressed as $e^{\Xo} e^{\Yo}= e^{\Zo}$ with
\beq\label{BCH0}
\Zo=\Xo+\Yo+\half[\Xo,\Yo]+\frac{1}{12}[\Xo,[\Xo,\Yo]]-\frac{1}{12}[\Yo,[\Xo,\Yo]]+\dots
\eeq
where the dots hide an infinite number of nested commutators of growing (eventually infinite) order. It is thus clear that when $\Xo$ and $\Yo$ commute to a c-number only the first three terms of the right hand side survive; otherwise the whole infinite series must be retained.

The General Ordering Theorem (GOT) proven in this paper establishes a connection between any pair of orderings that act on operators satisfying any commutation relation. 
As such, the GOT should be considered an ultimate result, as it is the most general ordering theorem that can be proven, both in terms of orderings and of operators. %The theorem itself as well as its proof are rigorous, but purely mathematical issues (like e.g. boundedness of the operators and convergence of the series) are not treated as they go far beyond the scope of this Letter (and of the author's skills). As exculpatory evidence for this, we mention that convergence of the BCH formula has been an intensive research topic for about ... years~\cite{}.

\section{Setup} Let $\phio=\{\phio_\alpha\}$ be a set of operators with $\alpha\in\Omega$, the index set. Let $\Oc$ be an ordering, which rearranges any product of operators $\phio_1\dots\phio_n$ according to the ordering rule $p_1\succ\dots \succ p_n$:
 \beq\label{Ord}
\Oc[\phio_1\dots\phio_n]= \phio_{p_1}\dots\phio_{p_n}\,,
\eeq
where  $p_1\dots p_n$ is a permutation of the input string $1\dots n$. 
%There exist orderings that display a different signature, depending on the number of permutations needed to obtain the output string from the input. In this Letter we are not interested to these orderings for the simple reason that they are of little (if any) physical interest~\cite{FerDio21}. 
Orderings of this type are called {\it monomials} because the ordered output is a single permutation of the input string. Typical examples of monomial orderings are the time ordering ($\Tc$), which orders the integers increasingly from right to left,
%\beq
%\Tc[\phio_1\dots\phio_n]= \phio_{n}\dots\phio_{1}\,;
%\eeq
and the normal ordering ($\Nc$), which pushes all creation operators to the left of all annihilation operators.
%\beq
%\Nc[a^\dag a a^\dag a^\dag]= (a^\dag)^3 a\,.
%\eeq
 The orderings where the output is a weighted sum of permutations are called {\it non-monomials}~\cite{Dio18}:
 \beq\label{nmOrd}
\Oc[\phio_1\dots\phio_n]=\sum_P w_P [\phio_{p_1}\dots\phio_{p_n}]_P\,,
\eeq
where $P$ denotes a specific permutation. Since the idea underlying orderings is that they simply rearrange the input string without changing the multiplicity of the input operators, the weights $w_P$ must sum to one. A notorious example of non-monomial ordering is the Weyl ordering ($\Wc$), which outputs the fully symmetrized version of the input product, 
%\beq
%\Wc[\phio_1\phio_2]=\half(\phio_1\phio_2+\phio_2\phio_1)\,,
%\eeq
or which can be equivalently defined as the identity on exponentials with linear arguments:
\beq\label{Weyl}
\Wc\left[e^{a\phio_1+b\phio_2}\right]=e^{a\phio_1+b\phio_2}\,,
\eeq
(because these are inherently symmetric).
The GOT applies to all monomial orderings, and to those non-monomial orderings that can be expressed in terms of a monomial one. In~\cite{Dio18} it was shown that $\Wc$ belongs to the latter class, as it can be expressed in terms of $\Tc$.

We assume that the operators $\phio_\alpha$
are linear combinations of operators in $\vphio=\{\vphio_k\}$, $k\in\Omega'$:
\beq\label{phitov}
\phio_\alpha=\Lc_{\alpha k} \vphio_k\,, 
\eeq
where the Einstein sum convention for repeated indexes is assumed, with the condition that sums run on the elements of the index set respective to the repeated index. 
We thus introduce a different ordering $\Oc'$, which orders the operators $\vphio_k$ according to the rule $k_1\Yright\dots\Yright k_n$, but may not order the operators $\phio_\alpha$. 
Similarly, the operators $\vphio_k$ may not be ordered by $\Oc$. 
The decomposition~\eqref{phitov} allows us to define indirectly the $\Oc'$-ordering of operators $\phio_\alpha$ as follows:
\beq\label{Ordpr}
\Oc'\left[\prod_{i=1}^n\phio_{\alpha_i}\right]\equiv
\Oc'\left[\prod_{i=1}^n\Lc_{\alpha_ik_i}\vphio_{k_i}\right]\,.
\eeq
In the following we assume $\Omega$ and $\Omega'$ to be discrete sets, but the results obtained hold invariably for continuous indexes, provided that sums are upgraded to intergrals, and partial derivatives become functional derivatives.

There are two key ingredients in the GOT. The first is the concept of contraction, that is the difference of how $\Oc$ and $\Oc'$ order a pair of operators of $\phio$:
\beq\label{Contr}
\hC_{\alpha\beta}\equiv(\Oc-\Oc')\phio_\alpha\phio_\beta
=(\theta_{l\Yright k}-\theta_{\beta\succ\alpha})\Lc_{\alpha k}\Lc_{\beta l}[\vphio_k,\vphio_l]\,,
\eeq
where $\theta_{\beta\succ\alpha}=1$ if $\beta\succ\alpha$, and $\theta_{l\Yright k}=1$ if $l\Yright k$, both zero otherwise. % where the prime is only meant to recall that this theta reflects the ordering rule associated to $\Oc'$. 
We remark that, here and in the forthcoming, such unit step functions $\theta$ are excluded from the Einstein notation; it is thus understood that in Eq.~\eqref{Contr} only the sums over $k,l$ are performed (see Appendix A).
Similarly, we can define the contraction for the operators in $\vphio$:
\beq\label{Contrv}
\hcomp_{kl\alpha\beta} \equiv(\Oc-\Oc')\vphio_k\vphio_l
=(\theta_{l\Yright k}-\theta_{\beta\succ\alpha})[\vphio_k,\vphio_l]\,
\eeq
which satisfies $\hC_{\alpha\beta}=\Lc_{\alpha k}\Lc_{\beta l}\hcomp_{kl\alpha\beta}$. We remark that $\hC$ is a symmetric matrix ($\hC_{\alpha\beta}=\hC_{\beta\alpha}$), while the elements of $\hcomp$ satisfy $\hcomp_{kl\alpha\beta}=\hcomp_{lk\beta\alpha}$ and $\hcomp_{kk\alpha\beta}=0$. %In order to improve the readability of this paper we use the simplified notation of Eq.~\eqref{Contrv}, and we restore all indexes where necessary. 

The second ingredient is a differential operator which has the peculiarity of replacing the differentiated operator by another one:
\beq\label{tens}
\left(\hat X\bdot\frac{\partial}{\partial\hat B}\right)\hat A\hat B\hat C=\hat A\hat X\hat C\,.
\eeq
In the modern jargon, such a mathematical object is called {\it tensor (or operator) directional derivative} (with respect to $\hat B$, in the direction $\hat X$)~\cite{Its07,Kel22}, and it belongs to the family of G\^ateaux derivatives~\cite{Spi71}. We should nonetheless mention that this differential operator dates back at least to the early proofs of the BCH by Baker~\cite{Bak05} and Hausdorff~\cite{Hau06}, who respectively called it ``substitutional'' and ``polar'' derivative. This is a linear operator that satisfies suitable Leibniz and chain rules, and that behaves like a standard partial derivative when $X$ is a c-number (see Appendix A for further details).  In order to keep the notation contained, we will use the following shorthand notation: $\hp_\alpha\equiv\partial/\partial\phio_\alpha$ and
$\hud_k\equiv\partial/\partial\vphio_k$. We are now ready to state the theorem.

\section{General Ordering Theorem} 
{\it Theorem.} Let $F(\phio)$ be a functional of operators in the set $\phio=\{\phio_\alpha\}$, $\alpha\in\Omega$, and let $\Oc$ be an ordering on $\phio$. Let equation~\eqref{phitov} hold for a set of operators $\vphio=\{\vphio_k\}$, $k\in\Omega'$, and let $\Oc'$ be an ordering on this set. Let $\succ$ and $\Yright$ denote the ordering rules of $\Oc$ and $\Oc'$ respectively. Then,
\beq\label{GOT1}
\mathcal{O}[F(\phio)]=\mathcal{O}'[F(\phio')]\,,
\eeq
with $\phio'=\{\phio'_\alpha\}$, $\alpha\in\Omega$, and
\beqa
\label{phiop} \phio'_\alpha&\equiv&\phio_\alpha+\hat{\Cc}_{\alpha\beta}\bdot\hp_\beta\,.
\eeqa

\begin{proof}
%{\it Proof.} 
In order to keep the treatment as simple as possible, we report here the proof for the special case where $\Lc_{\alpha k}=\delta_{\alpha k}$ (Kronecker delta), which implies $\{\phio_\alpha\}=\{\vphio_k\}$ and $\Omega=\Omega'$, i.e. $\Oc$ and $\Oc'$ order the same operators but according to different rules (like e.g. normal and anti-normal orderings).  In such a case Eq.~\eqref{Contr} simplifies to
\beq\label{Contr2}
\hC_{\alpha\beta}=(\theta_{\beta\Yright\alpha}-\theta_{\beta\succ\alpha})[\phio_\alpha,\phio_\beta]\,.
\eeq 
Some technical details of this proof are left to Appendix B, while the general proof for $\Oc$ and $\Oc'$ ordering different sets of operators is reported in Appendix C. We rely on the power series expansion that any functional admits, and we prove the theorem by induction: we assume that the identity
\beq\label{nGOT}
\Oc\left[\prod_{i=1}^n\phio_{\alpha_i}\right]=\Oc'\left[\prod_{i=1}^n\phio'_{\alpha_i}\right]
\eeq
holds up to an arbitrary $n$, and we prove that 
\beq\label{n+1GOT}
\Oc\left[\phio_{\alpha}\prod_{i=1}^n\phio_{\alpha_i}\right]=\Oc'\left[\phio'_{\alpha}\prod_{i=1}^n\phio'_{\alpha_i}\right]\,.
\eeq
We implicitly assume that $\Oc'$ orders the operators $\phio'_\alpha$ {\it before} Eq.~\eqref{phiop} is replaced. In other words, if we define the ``priming operator'' $\mathbb{P}$ as $\mathbb{P}[\phio_\alpha]\equiv\phio'_\alpha$, then
\beq
\Oc'\left[\prod_{i=1}^n\phio'_{\alpha_i}\right]\equiv\mathbb{P}\left\{\Oc'\left[\prod_{i=1}^n\phio_{\alpha_i}\right]\right\}\,.
\eeq
Accordingly, whether $\hC_{\alpha\beta}$ belongs or not in $\Omega'$ is irrelevant.
We also remark that the product of $n$ operators of $\phio'$ always contains $n$ operators of $\phio$, and its rightmost operator is not primed because its derivates have nothing to act upon on their right.
%Accordingly, whether $\hC_{\alpha\beta}$ does or does not belong to $\Omega'$ is irrelevant.
Without loss of generality, we assume that $\Oc$ orders the operators of $\phio$ according to the rule $\alpha_n\succ\dots\succ\alpha_{j+1}\succ\alpha\succ\alpha_j\succ\dots\succ\alpha_1$, which implies that
\beq
%\label{On} \Oc\left[\prod_{i=1}^n\phio_{\alpha_i}\right]&=&\phio_{\alpha_n}\dots\phio_{\alpha_1}\\
\label{On+1} \Oc\left[\phio_{\alpha}\prod_{i=1}^n\phio_{\alpha_i}\right]=\phio_{\alpha_n}\dots\phio_{\alpha_{j+1}}\phio_{\alpha}\phio_{\alpha_j}\dots\phio_{\alpha_1}\,.
\eeq
This equation can be rewritten in terms of the left hand side of the hypothesis~\eqref{nGOT} as follows
\beqa\label{leftpol}
\Oc\left[\phio_{\alpha}\prod_{i=1}^n\phio_{\alpha_i}\right]=\left(\phio_{\alpha}- \theta_{\beta\succ\alpha}[\phio_{\alpha},\phio_\beta]\boldsymbol{\cdot}\hat{\partial}_{\beta}\right)\mathcal{O}\left[\prod_{i=1}^n\phio_{\alpha_i}\right]\,,
\eeqa
(see the Appendix B for the proof), which in turn reads
\beqa\label{leftpol2}
\Oc\left[\phio_{\alpha}\prod_{i=1}^n\phio_{\alpha_i}\right]=\left(\phio'_{\alpha}- \theta_{\beta\Yright\alpha}[\phio_{\alpha},\phio_\beta]\boldsymbol{\cdot}\hat{\partial}_{\beta}\right)\mathcal{O}'\left[\prod_{i=1}^n\phio'_{\alpha_i}\right],
\eeqa
after Eqs.~\eqref{phiop}-\eqref{nGOT} are used. The first term on the right hand side can be rewritten by exploiting an identity similar to Eq.~\eqref{leftpol} (with $\phio$ replaced by $\phio'$, and $\Oc$ by $\Oc'$ - see Appendix B), which finally gives
\beqa\label{leftpol3}
\Oc\left[\phio_{\alpha}\prod_{i=1}^n\phio_{\alpha_i}\right]&=&\Oc'\left[\phio'_{\alpha}\prod_{i=1}^n\phio'_{\alpha_i}\right]\nonumber\\
&&\hspace{-2.5cm}+\theta_{\beta\Yright\alpha}\left([\phio'_{\alpha},\phio'_\beta]\boldsymbol{\cdot}\hp'_\beta-[\phio_{\alpha},\phio_\beta]\boldsymbol{\cdot}\hat{\partial}_{\beta}\right)\mathcal{O}'\left[\prod_{i=1}^n\phio'_{\alpha_i}\right],\hspace{0.5cm}
\eeqa
where $\hp'_\beta=\partial/\partial\phio'_\beta$. The GOT is thus proven if the operators
\beq\label{Dop}
\Dc_\alpha=\theta_{\beta\Yright\alpha}\left([\phio'_{\alpha},\phio'_\beta]\boldsymbol{\cdot}\hp'_\beta-[\phio_{\alpha},\phio_\beta]\boldsymbol{\cdot}\hat{\partial}_{\beta}\right)
\eeq
are identically zero. In order to prove it, we exploit the definition~\eqref{phiop} on the first term of $\Dc_\alpha$, obtaining
\beq\label{commp}
[\phio'_{\alpha},\phio'_\beta]=[\phio_{\alpha},\phio_\beta]+[\hC_{\alpha\gamma}\bdot\hp_\gamma,\hC_{\beta\varepsilon}\bdot\hp_\varepsilon]\,,
\eeq
while the second term is rewritten by exploiting the chain rule for tensorial derivatives~\cite{Its07,Kel22}:
\beqa\label{chain}
\theta_{\beta\Yright\alpha}[\phio_{\alpha},\phio_\beta]\boldsymbol{\cdot}\hat{\partial}_{\beta}&=&\theta_{\beta\Yright\alpha}\left([\theta_{\gamma\Yright\alpha}[\phio_{\alpha},\phio_\gamma]\bdot\hp_\gamma,\hC_{\beta\varepsilon}\bdot\hp_\varepsilon]\right.\nonumber\\
&&\hspace{1cm}\left.+[\phio_{\alpha},\phio_\beta]\right)\boldsymbol{\cdot}\hat{\partial}'_{\beta}.
\eeqa
By replacing these equations in Eq.~\eqref{Dop} one is left with
\beq\label{Dfin}
\Dc_\alpha= \theta_{\beta\Yright\alpha}[\theta_{\gamma\succ\alpha}[\phio_{\alpha},\phio_\gamma]\bdot\hp_\gamma,\hC_{\beta\varepsilon}\bdot\hp_\varepsilon]\boldsymbol{\cdot}\hat{\partial}'_{\beta}\,,
\eeq
which with a simple calculation can be shown to be identically zero (see Appendix B).
Equation~\eqref{leftpol3} thus reduces to Eq.~\eqref{n+1GOT}, which completes the induction and, with it, our proof of the GOT. %\qedsymbol
\end{proof}

We remark that the GOT equally applies when Eq.~\eqref{phitov} is replaced by the more general linear relationship
%\beq
$\lambda_\alpha\phio_\alpha=\uplambda_k\phio_k$.
%\eeq
The price one has to pay is that the functional to be ordered must display the dependence $F(\lambda_\alpha\phio_\alpha)$, because only in this case a decomposition in terms of $F(\uplambda_k\vphio_k)$ is allowed.

\section{GOT at work} A first formal check can be done by considering operators with c-number contraction (free bosonic fields, canonical operators, etc.). In this case the differential operator in Eq.~\eqref{tens} reduces to a standard partial derivative, and the GOT recovers the ordering theorem for bosons derived in~\cite{FerDio21}. Needless to say, when one further specializes to time and normal orderings, Wick's theorem is also recovered.
We now elucidate how the GOT works with operators that possess operatorial commutation relations, by considering few simple examples.

As a warmup we consider $\Oc=\Tc$, the time ordering defined in the introduction ($n\succ\dots\succ1$), while $\Oc'$ is the respective anti-ordering $\overline{\Tc}$ ($1\Yright\dots\Yright n$). %which orders increasingly from left to right). 
Assume that we want to rewrite the time ordered version of the product $\phio_1\phio_2\phio_3$ as an anti-time ordered product. According to Eq~\eqref{Contr2} the entries of the contraction matrix are given by: $\hC_{12}=(\theta_{2\Yright1}-\theta_{2\succ1})[\phio_1,\phio_2]=-[\phio_1,\phio_2]$, $\hC_{13}=-[\phio_1,\phio_3]$, and $\hC_{23}=-[\phio_2,\phio_3]$. The GOT prescribes that
\beq\label{TantiT}
\Tc\left[\phio_1\phio_2\phio_3\right]=\phio_3\phio_2\phio_1=\overline{\Tc}[\phio'_1\phio'_2\phio'_3]=\phio'_1\phio'_2\phio_3\,,
\eeq
where 
\beqa
\phio'_1&=&\phio_1-[\phio_1,\phio_2]\bdot\hp_2-[\phio_1,\phio_3]\bdot\hp_3\\
\phio'_2&=&\phio_2+[\phio_2,\phio_1]\bdot\hp_1-[\phio_2,\phio_3]\bdot\hp_3\,,
\eeqa
and $\phio_3$ is not primed because, as previously mentioned, the derivatives of rightmost term in any product have nothing to act upon.
By replacing these two equations in the rightmost term of Eq.~\eqref{TantiT} we find
\beqa
\phio'_1\phio'_2\phio_3\!\!\!&=&\!\!\!\left(\!\phio_1-\!\![\phio_1,\phio_2]\!\bdot\!\hp_2\!-\![\phio_1,\phio_3]\!\bdot\!\hp_3\!\right)\!\!\left(\!\phio_2\phio_3\!-\![\phio_2,\phio_3]\!\right)\nonumber\\
&=&\!\!\!\phio_1\phio_3\phio_2\!-\!\phio_3[\phio_1,\phio_2]\!-\![\phio_1,\phio_3]\phio_2=\phio_3\phio_2\phio_1\,,
\eeqa
which confirms the correctness of the GOT.

We now upgrade to an example where $\Oc$ and $\Oc'$ order different sets of operators. Let $\Oc=\mathcal{A}$, which orders the indexes in alphabetical order from right to left ($Z\succ\dots\succ A$), and $\Oc'=\Tc$ ($n\Yright\dots\Yright 1$). Moreover, let Eq.~\eqref{phitov} hold, with $\Omega=\{A,B\}$ and $\Omega'=\{1,2\}$:
\beqa
\phio_A&=&\Lc_{A1}\vphio_1+\Lc_{A2}\vphio_2\\
\phio_B&=&\Lc_{B1}\vphio_1+\Lc_{B2}\vphio_2\,.
\eeqa
We aim at expressing the alphabetically ordered product of the operators $\phio_A,\phio_B$ as a time ordered product. According to the GOT, the following identity holds:
\beqa
\mathcal{A}[\phio_A\phio_B]&=&\Tc[\phio'_A\phio'_B]\nonumber\\
&=&\Tc[(\Lc_{A1}\vphio'_1+\Lc_{A2}\vphio'_2)(\Lc_{B1}\vphio'_1+\Lc_{B2}\vphio'_2)]\hspace{0.4cm}
\eeqa
where in the second line we have decomposed the operators of $\phio'$ in terms of $\vphio'$ in order to be able to apply $\mathcal{T}$ (see Appendix C). By explicitating the orderings one finds
\beqa\label{AvT}
\phio_B\phio_A&=&\Lc_{A1}\Lc_{B1}\vphio_1\vphio_1+\Lc_{A2}\Lc_{B2}\vphio_2\vphio_2\nonumber\\
&&+\Lc_{A1}\Lc_{B2}\vphio'_2\vphio_1+\Lc_{A2}\Lc_{B1}\vphio'_2\vphio_1\,,
\eeqa
with
\beqa\label{AvT2}
\vphio'_2&=&\vphio_2+(\theta_{1\Yright2}-\theta_{\beta\succ\alpha})[\vphio_2,\vphio_1]\bdot\hud_1\,.
\eeqa
In the first line of Eq.~\eqref{AvT} the primes of the leftmost operators disappeared because $\hcomp_{kk\alpha\beta}=0$, while the primes of the rightmost operators of both lines dropped because their derivatives have nothing to act upon.
We observe that the differential part of Eq.~\eqref{AvT2} depends on the values that $\alpha$ and $\beta$ take in $\Omega$ (see Appendix C). In the term $\Lc_{A1}\Lc_{B2}\vphio'_2\vphio_1$ of Eq.~\eqref{AvT}, the indexes 1 and 2 are associated respectively to $A$ and $B$ (as recalled by the coefficients $\Lc$) because the operator $\vphio_1$ comes from the decomposition of $\phio_A$, while the operator $\vphio_2$ comes from the decomposition of $\phio_B$. This implies that in the contraction displayed by Eq.~\eqref{AvT2} one has $\beta=A$ and $\alpha=B$. In the term $\Lc_{A2}\Lc_{B1}\vphio'_2\vphio_1$ the association is the opposite, which implies $\beta=B$ and $\alpha=A$. Since $\theta_{1\Yright2}=0$, $\theta_{A\succ B}=0$ and $\theta_{B\succ A}=1$, one easily finds that
\beqa\label{AvT3}
\Lc_{A1}\Lc_{B2}\vphio'_2\vphio_1&=&\Lc_{A1}\Lc_{B2}\vphio_2\vphio_1\\
\Lc_{A2}\Lc_{B1}\vphio'_2\vphio_1&=&\Lc_{A2}\Lc_{B1}(\vphio_2-[\vphio_2,\vphio_1]\bdot\hud_1)\vphio_1\nonumber\\
&=&\Lc_{A2}\Lc_{B1}\vphio_1\vphio_2\,,
\eeqa
which once replaced in Eq.~\eqref{AvT} make it trivial to check that the identity holds.
Now that the basic working principles of the GOT are clear, we move to more complicated applications which disclose its vast potential.

\subsection{Baker-Campbell-Hausdorff Formula} 
This formula originated in the context of Lie groups, where it is a natural question to ask if the products of two group transformations $e^{\Xo}e^{\Yo}$ can be expressed as a single exponential~\cite{Cam98,Poi00}. What physicists typically learn about the BCH is the following: if $\Xo$ and $\Yo$ commute to a c-number the BCH is a very helpful tool because it displays a nice and compact structure; otherwise it is a disaster. In mathematical terms this is translated as follows: 
\beq\label{BCH1}
e^{\Xo} e^{\Yo}= e^{\Zo}
\eeq
 with
\beqa
&&\Zo=\sum_{n=0}^\infty\frac{1}{n!}\left[W(\Xo,\Yo)\bdot\hp_Y\right]^n\Yo\,.\\
&&W(\Xo,\Yo)=\sum_{n=0}^\infty \frac{B_n}{n!} {\rm ad}_{\Yo}^n\Xo\,,\label{W}
\eeqa
where $B_n$ are the Bernoulli numbers, ${\rm ad}_{\Yo}\Xo=[\Yo,\Xo]$ and ${\rm ad}_{\Yo}^n\Xo=[\Yo,{\rm ad}_{\Yo}^{n-1}\Xo]$.
This is an infinite series involving nested commutators of growing (eventually infinite) order, whose terms up to order three are displayed by Eq.~\eqref{BCH0}.
We now show that the GOT allows to express the BCH in a very elegant way. We introduce the ordering $\Nc_{XY}$, which is a sort of normal ordering that pushes all operators $\Xo$ to the left of the operators $\Yo$. It is a trivial task to show that~\cite{Dio18}
\beq
\Nc_{XY}\left[e^{\Xo+\Yo}\right]=e^{\Xo} e^{\Yo}\,.
\eeq
By recalling that the Weyl ordering acts as the identity on the exponential function (see Eq.~\eqref{Weyl}), we understand the BCH~\eqref{BCH1} as relation between $\Nc_{XY}$ and $\Wc$:
\beq
\Nc_{XY}\left[e^{\Xo+\Yo}\right]=\Wc\left[e^{\Zo}\right]\,.
\eeq
%\beq
%\Nc_{XY}\left[e^{\Xo+\Yo}\right]=\Wc\left[e^{\Zo}\right]\,.
%\eeq
Remarkably, a relation between these orderings can be established  also via the GOT, which states
\beq\label{aBCH}
\Nc_{XY}\left[e^{\Xo+\Yo}\right]=\Wc\left[e^{\Xo'+\Yo'}\right]\,,
\eeq
with
\beqa\label{BCHp}
\Xo'=\Xo,\qquad\Yo'=\Yo+[\Xo,\Yo]\bdot\hp_X\,,
\eeqa
that is
\beq\label{BCHGOT}
e^{\Xo} e^{\Yo}= e^{\Xo+\Yo+[\Xo,\Yo]\bdot\hp_X}\,.
\eeq
The expressions for the primed operators in Eq.~\eqref{BCHp} are dictated by the GOT, and they are derived in Appendix D.
We thus see that the GOT provides a very compact expression of the BCH, though being completely equivalent to it. Some comments are at the order. First, one may notice that the contraction in Eq.~\eqref{BCHp} is not symmetric ($\hC_{XY}\neq\hC_{YX}$). This is due to the non-monomial nature of $\Wc$, as explained in Appendix D.
%This is due to the fact that the Weyl ordering is non-monomial, while the GOT applies to monomial ordering only. However, there exists an equivalent expression of $\Wc$ in terms $\Tc$ [...], and is thus such an expression that is used to compute the contraction [...].
Second, one may be puzzled by the fact that although Eqs.~\eqref{BCH1}-\eqref{W} and Eq.~\eqref{BCHGOT} use the same mathematical ingredients, the latter displays such a simpler structure. The reason is that Baker and Hausdorff were looking for a closed expression explicitly displaying the elements of the Lie algebra (i.e. nested commutators)~\cite{Bak05,Hau06,AchBon12}. The GOT instead provides an ``open'' expression, in the sense that the nested commutators are obtained by applying it iteratively the directional derivative. 
We further observe that, by expanding in Taylor series the right hand side of Eq.~\eqref{BCHGOT}, the $n$-th order of expansion $\hat z_n(\Xo,\Yo)$ can be determined recursively via the simple formula
\beq\label{recBCH}
\hat z_n(\Xo,\Yo)=\frac{1}{n}(\Xo+\Yo+[\Xo,\Yo]\bdot\hp_X)\,\hat z_{n-1}(\Xo,\Yo)
\eeq
As a check of the correctness of the GOT, let us compute the third order term $3!\,\hat z_3(\Xo,\Yo)$, which amounts to
\beqa
&&\hspace{-0.3cm}(\Xo+\Yo+[\Xo,\Yo]\bdot\hp_X)(\Xo+\Yo+[\Xo,\Yo]\bdot\hp_X)(\Xo+\Yo)=\nonumber\\
&&\hspace{-0.3cm}=(\Xo+\Yo+[\Xo,\Yo]\bdot\hp_X)\left[(\Xo+\Yo)(\Xo+\Yo)+[\Xo,\Yo]\right]\nonumber\\
&&\hspace{-0.3cm}=\!(\!\Xo\!+\!\Yo\!)^3\!+\![\Xo\!,\!\Yo](\!\Xo\!+\!\Yo\!)\!+\!2(\!\Xo\!+\!\Yo)[\Xo\!,\!\Yo]\!+\![[\Xo\!,\!\Yo],\!\Yo].
\eeqa
It is again an elementary operation to check that this equation coincides with the third order expansion (i.e. involving products three operators) of $e^{\Zo}$, with $\Zo$ given in Eq.~\eqref{BCH0}.

\subsection{Magnus Expansion} W. Magnus looked for the solution of the initial value problem associated to the following ordinary differential equation:
\beq\label{evoeq}
\frac{d\hat{U}_t}{dt}=\hat A_t\hat U_t,\quad \hat U_0=\hat I
\eeq
where $\hat U_t$ and $\hat A_t$ are linear operators, and $\hat I$ is the identity. Precisely, Magnus was interested in a pure exponential solution of the type
\beq\label{ME}
\hat U_t= e^{\hat V_t}\,,
\eeq
and he found that $\hat V_t$ satisfies~\cite{Mag54}
\beq\label{MEser}
\frac{d\hat V_t}{dt}=\sum_{n=0}^\infty \frac{B_n}{n!} {\rm ad}_{\hat V}^n\Ao\,.
\eeq
Therefore, similarly to the BCH, also the ME consists of an infinite series of nested commutators of growing order (the ME is indeed also referred to as ``continuous BCH'').
Nonetheless, Eq.~\eqref{evoeq} is well known to physicists, as well as its solution that Dyson first wrote in terms of the time ordering operator:
\beq\label{evosol}
\hat U_t= \Tc \left[e^{\int_0^t \Ao_s ds}\right]\,.
\eeq
By recalling again Eq.~\eqref{Weyl}, we can understand the ME as the Weyl ordered version of Eq.~\eqref{evosol}:
\beq\label{}
 \Tc \left[e^{\int_0^t \Ao_sds}\right]=\Wc  \left[e^{\hat V_t}\right]\,.
\eeq
The GOT offers another way to relate time ordering to Weyl ordering, i.e. it provides a different expression for the ME, which reads
\beq\label{MEfin}
 \Tc \left[e^{\int_0^t \Ao_sds}\right]=e^{\int_0^t \left(\Ao_s+\int_0^t \theta_{us}[\Ao_u,\Ao_s]\bdot\hp_udu\right)ds}\,,
\eeq
where $\theta_{us}$ is the standard Heaviside step function. We eventually observe that a recursive formula similar to Eq.~\eqref{recBCH} holds also for the ME.
We leave the check of the validity of Eq.~\eqref{MEfin} at third order to Appendix D.

\section{Conclusions} We have proven the General Ordering Theorem, which establishes a connection between any pair of orderings acting on operators that posses generic commutation relations. After elucidating its working principles, we have demonstrated the potential of the GOT by establishing compact expressions both for the BCH and the ME. Besides being very elegant, these have the merit of providing simple recursive formulas to calculate efficiently high orders of expansion. This is not the case for the original versions of the BCH and ME, where the exponent contains infinite series, and in order to obtain the $n$-th order of expansion, one needs to put together and rearrange all lower orders.
Nonetheless, the GOT in some sense extends the range of applicability of the BCH and the ME: while the latter concern exponentials, the former applies to any functional.

On the more applied side, we mention few research fields which will benefit from the GOT. The first are non-Markovian open dynamics for spin systems (e.g. spin-boson and Jaynes-Cummings)~\cite{BrePet02}. 
Since in the time evolution of these systems the fermionic operators enter linearly, one cannot exploit the fermionic Wick's theorem, and the GOT is needed~\cite{FerDio21,Fer17}. Another field of application is nonlinear optics, whose processes (like e.g. spontaneous parametric down-conversion and four-wave mixing) involve photonic operators that do not commute to c-numbers (because of the nonlinearity), thus preventing the use of standard ordering theorems and requiring different techniques~\cite{Que14,Que15}. 
Concerning quantum field theory, one may exploit the GOT to express known quantities in terms of the ``preferred ordering'' introduced by Dalibard, J. Dupont-Roc, C. Cohen-Tannoudji~\cite{Daletal82}. Last, in the context of Yang-Mills theories, the GOT may be exploited to rewrite standard amplitudes in terms of color-ordered ones~\cite{Sch14}. 
We conclude by observing that, in general, any research topic where operator ordering or non-commutative algebra play a central role will potentially benefit from the GOT.

%keldysh? schwinger map

\section*{Acknowledgements} I wish to thank L. Di\'osi for endless stimulating discussions and continuous inspiration. I acknowledge M. Palma, F. Ciccarello and G. Gasbarri for useful discussions. Acknowledgement is also due for financial support from MUR through project PRIN (Project No. 2017SRN-BRK QUSHIP).

\appendix

\begin{widetext}
\section{Mathematical preliminaries}
In this paper we make extensive use of the Einstein notation for repeated indexes, with the exception of the unit step functions $\theta_{\alpha\succ\beta}$ and $\theta_{l\Yright k}$, whose indexes do not contribute to the repetition, and are only meant to limit the sums generated by the repetition of other indexes. Two illustrative examples are the definitions of contraction in Eq.~\eqref{Contr} and Eq.~\eqref{Contr2}. In Eq.~\eqref{Contr}, indexes $k,l$ are repeated because they are displayed both by $\Lc$ and by $\vphio$, while $\alpha,\beta$ are not repeated because $\theta_{\alpha\succ\beta}$ does not contribute to the repetition. Therefore, one should understand Eq.~\eqref{Contr} as
\beq\label{}
\hC_{\alpha\beta}=\sum_{k,l\in\Omega'}
(\theta_{l\Yright k}-\theta_{\beta\succ\alpha})\Lc_{\alpha k}\Lc_{\beta l}[\vphio_k,\vphio_l]\,.
\eeq
In Eq.~\eqref{Contr2} instead there is no sum, because $\theta_{\alpha\succ\beta}$ does not contribute to the repetition, thus explaining why the left hand side displays two indexes.
Excepting those belonging to unit step functions, all other repeated indexes are summed.

One of the two main ingredients of the GOT is a tensor (or operator) derivative, whose properties are here briefly reviewed; for further (and more formal) details we refer the reader to~\cite{Kel22,Spi71}. We start by recalling that the action of any differential operator on operator functionals relies on the Taylor expansion of the latter, making it sufficient to define the action of the former on a product of operators. An operator derivative $\hp_\alpha=\partial/\partial\phio_\alpha$ can be defined by
\beq
\hp_\alpha\left(\phio_n\dots\phio_\alpha\dots\phio_1\right)=\phio_n\dots\phio_1\,,
\eeq
and satisfies
\beq
[\hp_\alpha,\phio_\beta]=\delta_{\alpha\beta}\,,\qquad [\hp_\alpha,\hp_\beta]=0\,.
\eeq
The associated directional derivative is defined by
\beq
\phio_\beta\bdot\hp_\alpha\left(\phio_n\dots\phio_\alpha\dots\phio_1\right)=\phio_n\dots\phio_\beta\dots\phio_1\,,
\eeq
of which Eqs.~\eqref{tens} is a simplified example.
Given two functionals $F(\phio)$, $G(\phio)$ on $\phio=\{\phio_\alpha\}$, $\alpha\in\Omega$, the directional derivative defined above is a linear operator which satisfies product and chain rules:
\beqa
\phio_\beta\bdot\hp_\alpha\left[a\, F(\phio)+b\,G(\phio)\right]&=&a\,\phio_\beta\bdot\hp_\alpha F(\phio)+b\,\phio_\beta\bdot\hp_\alpha G(\phio)\qquad a,b\in\mathbb{C}\\
\phio_\beta\bdot\hp_\alpha \left[F(\phio)G(\phio)\right]&=&\left[\phio_\beta\bdot\hp_\alpha F(\phio)\right]G(\phio)+F(\phio)\left[\phio_\beta\bdot\hp_\alpha G(\phio)\right]\\
\phio_\beta\bdot\hp_\alpha F[G(\phio)]&=&\left[\phio_\beta\bdot\hp_\alpha G(\phio)\right]\bdot\frac{\partial F}{\partial G}\,.
\eeqa

\section{Details of the proof in the main text}
In this section we derive those identities that are used in the main text, but whose proofs were not reported there in order to keep the proof contained. We start by multiplying the left hand side of Eq.~(13) by $\phio_\alpha$, and then we move this operator towards right, with the aim of bringing it at the position that $\Oc$ assigns to it. After two switches one finds
\beqa\label{switchpol}
\phio_{\alpha}\Oc\left[\prod_{i=1}^n\phio_{\alpha_i}\right]&=&\phio_{\alpha}\phio_{\alpha_n}\dots\phio_{\alpha_1}=\phio_{\alpha_n}\phio_{\alpha}\phio_{\alpha_{n-1}}\dots\phio_{\alpha_1}+[\phio_{\alpha},\phio_{\alpha_n}]\phio_{\alpha_{n-1}}\dots\phio_{\alpha_1}\nonumber\\
&=&\phio_{\alpha_n}\phio_{\alpha_{n-1}}\phio_{\alpha}\phio_{\alpha_{n-2}}\dots\phio_{\alpha_1}+\phio_{\alpha_n}[\phio_{\alpha},\phio_{\alpha_{n-1}}]\phio_{\alpha_{n-2}}\dots\phio_{\alpha_1}+[\phio_{\alpha},\phio_{\alpha_n}]\phio_{\alpha_{n-1}}\dots\phio_{\alpha_1}\,,
\eeqa
where the second line can be rewritten as follows:
\beq
\phio_{\alpha}\Oc\left[\prod_{i=1}^n\phio_{\alpha_i}\right]=\phio_{\alpha_n}\phio_{\alpha_{n-1}}\phio_{\alpha}\phio_{\alpha_{n-2}}\dots\phio_{\alpha_1}+\left([\phio_{\alpha},\phio_{\alpha_{n}}]\bdot\hp_{\alpha_{n}}+[\phio_{\alpha},\phio_{\alpha_{n-1}}]\bdot\hp_{\alpha_{n-1}}\right)\Oc\left[\prod_{i=1}^n\phio_{\alpha_i}\right]\,.
\eeq
By iterating this procedure until $\phio_\alpha$ has reached the correct position, one finds
\beq
\phio_{\alpha}\Oc\left[\prod_{i=1}^n\phio_{\alpha_i}\right]=\Oc\left[\phio_{\alpha}\prod_{i=1}^n\phio_{\alpha_i}\right]+\theta_{\beta\succ\alpha}[\phio_{\alpha},\phio_\beta]\boldsymbol{\cdot}\hat{\partial}_{\beta}\mathcal{O}\left[\prod_{i=1}^n\phio_{\alpha_i}\right]\,,
\eeq
where $\theta_{\beta\succ\alpha}$ allows differentiation only of those terms that $\Oc$ places at the left of $\phio_\alpha$. This is Eq.~(17). Trivially, if we repeat the same calculation with $\phio'_\alpha$ and $\Oc'$, we find
\beq\label{staqua}
\phio'_{\alpha}\Oc'\left[\prod_{i=1}^n\phio'_{\alpha_i}\right]=\Oc'\left[\phio'_{\alpha}\prod_{i=1}^n\phio'_{\alpha_i}\right]+\theta_{\beta\Yright\alpha}[\phio'_{\alpha},\phio'_\beta]\boldsymbol{\cdot}\hat{\partial}'_{\beta}\mathcal{O}'\left[\prod_{i=1}^n\phio'_{\alpha_i}\right]\,,
\eeq
which is the identity used to pass from Eq.~(18) to Eq.~(19). The last step needed to complete the proof is to show that the operators $\Dc_\alpha$ in Eq.~(23) are identically zero. We first remark that, according to Eq.~(9), the derivative $\hp'_\beta$ in Eq.~(23) places the commutator at the position $\beta$, therefore the differential operator $\hp_\varepsilon$ inside the commutator can act only on those operators that $\Oc'$ places to the right of $\phio'_\beta$ ($\theta_{\beta\Yright\varepsilon}=1$). Accordingly, the term proportional to $\theta_{\varepsilon\Yright\beta}$ in $\hC_{\beta\varepsilon}$ vanishes, and one is left with
\beqa\label{Dlast}
\Dc_\alpha&=&- \theta_{\beta\Yright\alpha}\,\theta_{\gamma\succ\alpha}\theta_{\varepsilon\succ\beta}[[\phio_{\alpha},\phio_\gamma]\bdot\hp_\gamma,[\phio_{\beta},\phio_\varepsilon]\bdot\hp_\varepsilon]\boldsymbol{\cdot}\hat{\partial}'_{\beta}\nonumber\\
&=&-\theta_{\beta\Yright\alpha}\left(\theta_{\beta\succ\alpha}\theta_{\varepsilon\succ\beta}[[\phio_\alpha,\phio_\beta],\phio_\varepsilon]\bdot\hp_\varepsilon+\theta_{\varepsilon\succ\alpha}\theta_{\varepsilon\succ\beta}[\phio_\beta,[\phio_\alpha,\phio_\varepsilon]]\bdot\hp_\varepsilon-\theta_{\gamma\succ\alpha}\theta_{\alpha\succ\beta}[[\phio_\beta,\phio_\alpha],\phio_\gamma]\bdot\hp_\gamma\right.\nonumber\\
&&\hspace{1.5cm}\left.-\theta_{\gamma\succ\alpha}\theta_{\gamma\succ\beta}[\phio_\alpha,[\phio_\beta,\phio_\gamma]]\bdot\hp_\gamma\right)\boldsymbol{\cdot}\hat{\partial}'_{\beta}\nonumber\\
&=&-\theta_{\beta\Yright\alpha}\left(\theta_{\beta\succ\alpha}\theta_{\gamma\succ\beta}-\theta_{\gamma\succ\alpha}\theta_{\gamma\succ\beta}+\theta_{\gamma\succ\alpha}\theta_{\alpha\succ\beta}\right)\left([[\phio_\alpha,\phio_\beta],\phio_\gamma]\bdot\hp_\gamma\right)\boldsymbol{\cdot}\hat{\partial}'_{\beta}\,,
\eeqa
where the second line is obtained by performing the derivatives inside the commutator, and the third line by exploiting the Jacobi identity. %We then observe that, while the products $\theta_{\beta\alpha}\theta_{\gamma\beta}$ and $\theta_{\gamma\alpha}\theta_{\alpha\beta}$ are respectively associate to the ordered products $\gamma\beta\alpha$ and $\gamma\alpha\beta$.
We eventually observe that the product $\theta_{\gamma\succ\alpha}\theta_{\gamma\succ\beta}$ can be decomposed as $\theta_{\gamma\succ\alpha}\theta_{\gamma\succ\beta}=\theta_{\gamma\succ\beta}\theta_{\beta\succ\alpha}+\theta_{\gamma\succ\alpha}\theta_{\alpha\succ\beta}$, which replaced in Eq.~\eqref{Dlast} gives $\Dc_\alpha=0$. The GOT is proven.

\section{General proof}
This section is dedicated to the proof of the GOT in the general case where $\Oc$ and $\Oc'$ order different sets of operators, $\{\phio_\alpha\}$ and $\{\vphio_k\}$ respectively.
In analogy with the definition~\eqref{phiop} of the operators $\phio'_\alpha$, we introduce the operators $\vphio_k'$
\beq\label{vphipr}
\vphio_{k}'\equiv\vphio_k+\hcomp_{kl\alpha\beta}\bdot\hud_l\,,
\eeq
where $\hud_l=\partial/\partial\vphio_{l}$, which according to Eq.~(5) are such that
\beq\label{phiprtov}
\phio_\alpha'=\Lc_{\alpha k}\vphio_k'\,.
\eeq
The left hand side of Eq.~\eqref{vphipr} does not display the subscripts $\alpha,\beta$ because these do not play any role in the ordering of $\vphio_{k}'$ according to $\Oc'$. $\alpha,\beta$ can thus be dropped, provided that they are readily restored when the explicit expression for $\vphio_{k}'$ is used (see e.g. Eq.~\eqref{AvT2}).

%We observe that the left hand side of Eq.~\eqref{vphipr} does not display the subscripts $\alpha,\beta$ which are present at the right hand side. Indeed, $\alpha,\beta$ should not be considered as free parameters, because the are fixed by the operators $\phio'_\alpha$ and $\phio_\beta$ from whose decomposition $\vphio'_k$ and $\vphio_l$ are respectively obtained. In other words, $\alpha,\beta$  do not play any role in the ordering of $\vphio_{k}'$. We thus dropped these subscripts in the definition of $\vphio_{k}'$ in order to avoid convoluted notation, and restored them when the explicit expression for $\vphio_{k}'$ is used (see Eq.~\eqref{AvT2}).
%We also recall that the matrix  $\hcomp_{kl}$ satisfies $\Lc_{\alpha k}\Lc_{\beta l} \hcomp_{kl}=\hC_{\alpha\beta}$.
The general proof follows the one in the main text up to Eq.~(17) while, because of the general definition of contraction~(7), Eq.~(18) is replaced by
\beqa\label{leftpol4}
\Oc\left[\phio_{\alpha}\prod_{i=1}^n\phio_{\alpha_i}\right]=\left(\phio'_{\alpha}- \theta_{l\Yright k}\Lc_{\alpha k}[\vphio_k,\vphio_l]\boldsymbol{\cdot}\hud_{l}\right)\mathcal{O}'\left[\prod_{i=1}^n\phio'_{\alpha_i}\right],
\eeqa
where $\hud_l=\partial/\partial\vphio_{l}$, and we used $\hud_l=\Lc_{\beta l}\hp_\beta$.
 Before applying $\Oc'$ we need to decompose the operators $\phio'_\alpha$ in the first term of the right hand side according to Eq.~\eqref{phiprtov}:
\beq\label{firsttermpol}
\phio_{\alpha}'\,\Oc'\left[\prod_{i=1}^n\phio_{\alpha_i}'\right]
=\left(\Lc_{\alpha k}
\vphio_k'\right)\Oc' \left[\prod_{i=1}^n\left(\Lc_{\alpha_ik_i}\vphio_{k_i}'\right)\right]\,.                                                                                                                                      
\eeq
We introduce the identity corresponding to Eq.~\eqref{staqua} for the operators in $\vphio'$:
\beq\label{staqua2}
\vphio'_{k}\Oc'\left[\prod_{i=1}^n\vphio'_{k_i}\right]=\Oc'\left[\vphio'_{k}\prod_{i=1}^n\vphio'_{k_i}\right]+\theta_{l\Yright k}[\vphio'_{k},\vphio'_l]\boldsymbol{\cdot}\hud'_{l}\mathcal{O}'\left[\prod_{i=1}^n\vphio'_{k_i}\right]\,,
\eeq
which we exploit to rewrite the right hand side of Eq.~\eqref{firsttermpol} obtaining
\beq\label{}
\phio_{\alpha}'\,\Oc'\left[\prod_{i=1}^n\phio_{\alpha_i}'\right]
=\Oc' \left[\left(\Lc_{\alpha k}
\vphio_k'\right)\prod_{i=1}^n\left(\Lc_{\alpha_ik_i}\vphio_{k_i}'\right)\right]+\theta_{l\Yright k}\Lc_{\alpha k}[\vphio'_{k},\vphio'_l]\boldsymbol{\cdot}\hud'_{l}\Oc' \left[\prod_{i=1}^n\left(\Lc_{\alpha_ik_i}\vphio_{k_i}'\right)\right]\,                                                                                                                                      
\eeq
By replacing this equation into Eq.~\eqref{leftpol4} one eventually finds
\beq\label{leftpol5}
\Oc\left[\phio_{\alpha}\prod_{i=1}^n\phio_{\alpha_i}\right]=\Oc'\left[\phio'_{\alpha}\prod_{i=1}^n\phio'_{\alpha_i}\right]+\theta_{l\Yright k}\Lc_{\alpha k}\left([\vphio'_{k},\vphio'_l]\boldsymbol{\cdot}\hud'_{l}-[\vphio_{k},\vphio_l]\boldsymbol{\cdot}\hud_{l}\right)\Oc' \left[\prod_{i=1}^n\left(\Lc_{\alpha_ik_i}\vphio_{k_i}'\right)\right]\,.
\eeq
The differential operator in the second term of the right hand side has the same structure as $\Dc_\alpha$ in Eq.~(20), therefore one can prove that the former is identically zero by retracing the proof for $\Dc_\alpha=0$. This completes our proof.

\section{Contractions for the BCH and the ME}
As discussed in the main text, both the BCH and the ME can be understood as relation among orderings, one of which is the Weyl ordering. Accordingly, in order to apply the GOT, we first need to express the (non-monomial) Weyl ordering in terms of a monomial ordering. In~[41] it was shown that this can be done via the time ordering as follows:
\beq\label{WtoT}
\Wc\left[e^{a\phio_1+b\phio_2}\right]=\Tc\left[e^{\int_0^1 (a \phio_{1\tau}+b\phio_{2\tau})d\tau}\right]\,,
\eeq
where the time label $\tau$ is assigned only formally to the operators, i.e. $\phio_{1\tau}=\phio_{1}$ and $\phio_{2\tau}=\phio_{2}$, $\forall \tau$.
%In order to apply the GOT we thus have to exploit the right hand side of Eq.~\eqref{WtoT}, and then... 
We start with the BCH and we rewrite Eq.~(40) with the redundant notation of Eq.~\eqref{WtoT}:
\beq\label{}
\Nc_{XY}\left[e^{\int_0^1(\Xo_\tau+\Yo_\tau) d\tau}\right]=\Tc\left[e^{\int_0^1(\Xo'_\tau+\Yo'_\tau) d\tau}\right]\,.
\eeq
In order to write the explicit expressions for the primed operators, we need to compute the associated contractions, which read
\beqa
(\Nc_{XY}-\Tc)\Xo_\tau\Yo_\sigma&=&\Xo_\tau\Yo_\sigma-(\theta_{\tau\sigma}\Xo_\tau\Yo_\sigma+\theta_{\sigma\tau}\Yo_\sigma \Xo_\tau)=\theta_{\sigma\tau}[\Xo_\tau,\Yo_\sigma]\\
(\Nc_{XY}-\Tc)\Yo_\tau\Xo_\sigma&=&\Xo_\sigma\Yo_\tau-(\theta_{\tau\sigma}\Yo_\tau\Xo_\sigma+\theta_{\sigma\tau}\Xo_\sigma \Yo_\tau)=\theta_{\tau\sigma}[\Xo_\sigma,\Yo_\tau]\,,
\eeqa
where we dropped the symbol $\Yright$ because when the time ordering is involved $\theta_{\tau\Yright\sigma}$ coincides with the standard Heaviside step function $\theta_{\tau\sigma}$. Accordingly,
\beqa
\Xo'_\tau&=&\Xo_\tau+\theta_{\sigma\tau}[\Xo_\tau,\Yo_\sigma]\bdot\hp_{Y_{\sigma}}\\
\Yo'_\tau&=&\Yo_\tau+\theta_{\tau\sigma}[\Xo_\sigma,\Yo_\tau]\bdot\hp_{X_{\sigma}}\,.
\eeqa
We remark that these operators are placed by $\Tc$ at the position $\tau$, which implies that directional derivatives always act to the right of $\tau$, i.e. $\theta_{\sigma\tau}=0$ and $\theta_{\tau\sigma}=1$. Accordingly,
\beqa
\Xo'_\tau&=&\Xo_\tau\\
\Yo'_\tau&=&\Yo_\tau+[\Xo_\sigma,\Yo_\tau]\bdot\hp_{X_{\sigma}}\,,
\eeqa
which return Eq.~(41) once the original notation is restored.

Something similar occurs with the ME, that establishes a connection between time ordering and Weyl ordering, which according to the GOT can be expressed as follows:
\beq\label{GOTME}
 \Tc \left[e^{\int_0^t \Ao_sds}\right]=\Wc  \left[e^{\int_0^t \Ao'_sds}\right]\,.
\eeq
Since we need to exploit the decomposition~\eqref{WtoT}, we rename the time ordering in the left hand side of Eq.~\eqref{GOTME} as $\Tc\equiv\Tc_1$, while we call $\Tc_2$ the time ordering associated to $\Wc$ by Eq.~\eqref{WtoT}. We thus rewrite Eq.~\eqref{GOTME} as
\beq\label{71}
 \Tc_1 \left[e^{\int_0^1(\int_0^t \Ao_{s\sigma}ds)d\sigma}\right]=\Tc_2\left[e^{\int_0^1(\int_0^t \Ao'_{s\sigma}ds)d\sigma}\right]\,,
\eeq
where $\Tc_1$ orders the first subscript of $\Ao$, and $\Tc_2$ orders the second. The contraction in $\Ao'_{s\sigma}$ reads
\beqa
(\Tc_{1}-\Tc_2)\Ao_{s\sigma}\Ao_{u\upsilon}&=&\theta_{su}\Ao_{s\sigma}\Ao_{u\upsilon}+\theta_{us}\Ao_{u\upsilon}\Ao_{s\sigma}-\theta_{\sigma\upsilon}\Ao_{s\sigma}\Ao_{u\upsilon}-\theta_{\upsilon\sigma}\Ao_{u\upsilon}\Ao_{s\sigma}\nonumber\\
&=&(\theta_{su}-\theta_{\sigma\upsilon})[\Ao_{s\sigma},\Ao_{u\upsilon}]\,,
\eeqa
which leads to
\beq
\Ao'_{s\sigma}=\Ao_{s\sigma}+(\theta_{su}-\theta_{\sigma\upsilon})[\Ao_{s\sigma},\Ao_{u\upsilon}]\bdot\hp_{\Ao_{u\upsilon}}=\Ao_{s\sigma}+\theta_{us}[\Ao_{u\upsilon},\Ao_{s\sigma}]\bdot\hp_{\Ao_{u\upsilon}}\,,
\eeq
where the second identity is given  by the fact that $\upsilon$ always stands to the right of $\sigma$ (i.e. $\theta_{\sigma\upsilon}=1$). By replacing the result in Eq.~\eqref{71} and restoring the original notation, one eventually recovers Eq.~(50).
We now check the validity of the GOT by expanding Eq.~(50), whose third order reads:
\beqa
&&\hspace{-1.5cm}\frac{1}{3!}\left[\int_0^t \left(\Ao_{l}+\int_0^t \theta_{s_2l}[\Ao_{s_2},\Ao_{l}]\bdot\hp_{s_2}ds_2\right)dl\right]\left[\int_0^t \left(\Ao_{u}+\int_0^t \theta_{s_1u}[\Ao_{s_1},\Ao_{u}]\bdot\hp_{s_1}ds_1\right)du\right]\left[\int_0^t \Ao_sds\right]=\nonumber\\
&=&\frac{1}{3!}\left[\int_0^t \left(\Ao_{l}+\int_0^t \theta_{s_2l}[\Ao_{s_2},\Ao_{l}]\bdot\hp_{s_2}ds_2\right)dl\right]\left[\left(\int_0^t \int_0^t \Ao_u\Ao_s+\theta_{su}[\Ao_{s},\Ao_{u}]du\,ds\right)\right]\nonumber\\
&=&\int_0^t \int_0^t \int_0^t\bigg(\frac{1}{6}\Ao_{l}\Ao_u\Ao_s+\frac{1}{3} \theta_{su}\Ao_{l}[\Ao_{s},\Ao_{u}]+\frac{1}{6}\theta_{su}[\Ao_{s},\Ao_{u}]\Ao_{l}\nonumber\\
&&\hspace{2.5cm}+\frac{1}{6} \theta_{su}\theta_{sl}[[\Ao_{s},\Ao_{l}],\Ao_u]+\frac{1}{6}\theta_{su}\theta_{ul}[\Ao_{s},[\Ao_{u},\Ao_l]]\bigg)du\,ds\,dl\,.
\eeqa
By decomposing $\theta_{su}\theta_{sl}=\theta_{su}\theta_{ul}+\theta_{sl}\theta_{lu}$ and by rearranging the terms, we can rewrite the last identity as follows:
\beqa\label{MEex}
&=&\int_0^t \int_0^t \int_0^t\bigg(\frac{1}{6}\Ao_{l}\Ao_u\Ao_s+\frac{1}{4} \theta_{su}\Ao_{l}[\Ao_{s},\Ao_{u}]+\frac{1}{4}\theta_{su}[\Ao_{s},\Ao_{u}]\Ao_{l}+\frac{1}{6} \theta_{sl}\theta_{lu}[[\Ao_{s},\Ao_{l}],\Ao_u]+\frac{1}{6}\theta_{su}\theta_{ul}[\Ao_{s},[\Ao_{u},\Ao_l]]\nonumber\\
&&\hspace{2.5cm}+\frac{1}{6} \theta_{su}\theta_{ul}[[\Ao_{s},\Ao_{l}],\Ao_u]+\frac{1}{12}\theta_{su}[\Ao_{l},[\Ao_{s},\Ao_u]]\bigg)du\,ds\,dl\,,
\eeqa
where the second line can be shown to be identically zero by decomposing $\theta_{su}=\theta_{ls}\theta_{su}+\theta_{sl}\theta_{lu}+\theta_{su}\theta_{ul}$, and by exploiting the Jacobi identity.
On the other side, the first three terms of the Magnus series in Eq.~(47) read
\beqa
V(t)=\int_0^t \Ao_{s}ds+\half\int_0^t \int_0^t \theta_{su}[\Ao_{s},\Ao_{u}]+\frac{1}{6}\int_0^t \int_0^t \int_0^t \left(\theta_{sl}\theta_{lu}[[\Ao_{s},\Ao_{l}],\Ao_u]+\theta_{su}\theta_{ul}[\Ao_{s},[\Ao_{u},\Ao_l]]\right)du\,ds\,dl\,.
\eeqa
By expanding the exponential in Eq.~(46) and by retaining only the terms with three operators, one can easily check that the first line of~\eqref{MEex} is recovered.

\end{widetext}


\begin{thebibliography}{99}
\bibitem{Dys49} F.J. Dyson, Phys. Rev. {\bf 75}, 486 (1949).
\bibitem{Wic50} G.C. Wick, Phys. Rev. {\bf 80}, 268 (1950).
\bibitem{Wey27} H. Weyl, Z. Phys. {\bf 46}, 1 (1927).
\bibitem{Wig32} E.P. Wigner, Phys. Rev. {\bf 40}, 749 (1932).
\bibitem{CahGla69} K.E. Cahill and R.J. Glauber, Phys. Rev. {\bf 177}, 1857 (1969).
\bibitem{HouKin49} A. Hourlet and A. Kind, Helv. Phys. Acta {\bf 22}, 319 (1949).
\bibitem{And54} J. L. Anderson, Phys. Rev. {\bf94}, 703 (1954).
\bibitem{Mat55} T. Matsubara, Prog. Theor. Phys. {\bf 14}, 351 (1955).
\bibitem{Gau60} M. Gaudin, Nucl. Phys. {\bf 15}, 89 (1990).
\bibitem{Vagetal90} A. Vaglica , C. Leonardi, G. Vetri, J. Mod. Opt, {\bf 37}, 1487 (1990).
\bibitem{EvaSte96} T.S. Evans, D.A. Steer, Nucl. Phys. B {\bf 474}, 481 (1996).
\bibitem{Daletal82} J. Dalibard, J. Dupont-Roc, C. Cohen-Tannoudji, J. Physique {\bf 43}, 1617 (1982).
\bibitem{Kel65} L. V. Keldysh, Sov. Phys. JEPT {\bf20}, 1018 (1965).
\bibitem{Hal75} A. G. Hall, J. Phys. A: Math. Gen. {\bf8}, 214 (1975).
\bibitem{AgaWol69} G. S. Agarwal, E. Wolf, Lettere Nuovo Cimento,{\bf 1}, 140 (1969). 
\bibitem{AgaWol70} G. S. Agarwal, E. Wolf, Phys. Rev. D, {\bf 2}, 2206 (1970). 
\bibitem{AgaWol70b} G. S. Agarwal, E. Wolf, Phys. Rev. D, {\bf 2}, 2161 (1970). 
\bibitem{AgaWol70c} G. S. Agarwal, E. Wolf, Phys. Rev. D, {\bf 2}, 2187 (1970). 
\bibitem{PerCap77} J.H.H. Perk, H.W. Capel, Physica A {\bf89}, 265 (1977).
\bibitem{Peretal84} J.H.H. Perk, H.W. Capel, G.R.W. Quispel, and F.W. Nijhoff, Physica A {\bf123}, 1 (1984).
\bibitem{KutMuk97} W. Kutzelnigg, D. Mukherjee, J. Chem. Phys. {\bf 107}, 432 (1997).
\bibitem{Konetal10} L. Kong, M. Nooijen, D. Mukherjee, J. Chem. Phys. {\bf 132}, 234107 (2010).
\bibitem{Sch14} T. Schuster, Phys. Rev. D {\bf89}, 105022 (2014). 
\bibitem{SilPie82} B. Silvestre-Brac, R. Piepenbring, Phys. Rev. C {\bf26}, 2640 (1982).
\bibitem{Choetal85} K. Chou, Zh. Su, B. Hao, and L. Yu, Phys. Rep. {\bf 118} 1 (1985).
\bibitem{FerDio21} L. Ferialdi, L. Di\'osi, Phys. Rev. A {\bf104}, 052209 (2021). 
\bibitem{Mag50} W.Magnus, Ann. Math. {\bf52} 111 (1950).
%\bibitem{Mag66} W. Magnus, A. Karrass, and D. Solitar, {\it The Combinatorial Group Theory}. Wiley Interscience, Hoboken (1966).
\bibitem{Bor91} A. Borel, {\it Linear algebraic groups}, Graduate Texts in Mathematics {\bf 126}, Springer-Verlag (1991).
\bibitem{SchLyn01} P. E. Schupp, R. C. Lyndon, {\it Combinatorial group theory}, Springer-Verlag (2001).
\bibitem{FolSte82} G. B. Folland, and E.M. Stein, {\it Hardy spaces on homogeneous groups}. Mathematical Notes {28}, Princeton University Press (1982).
\bibitem{Hor67} L. H\"ormander, Acta Math. {\bf119}, 147 (1967).
\bibitem{Nagetal85} A. Nagel, E.M. Stein, and S. Wainger, Acta Math. {\bf 155}, 103 (1985).
\bibitem{Bou89} N. Bourbaki, {\it Lie Groups and Lie Algebras}, Springer-Verlag (1989).
\bibitem{Ser06} J.-P. Serre, {\it Lie Algebras and Lie Groups}, Springer (2006).
\bibitem{Sep07} M. R. Sepanski, {\it Compact Lie groups}. Graduate texts in mathematics {\bf235}, Springer (2007).
\bibitem{Mag54} W. Magnus, Comm. Pure and Appl. Math. {\bf7}, 649 (1954).
\bibitem{Blaetal} S. Blanes, F, Casas, J. A. Oteo, J. Ros, Phys. Rep. {\bf470}, 151 (2008).
\bibitem{Cam97} J. Campbell, Proc. Lond. Math. Soc. {\bf 28}, 381 (1897).
\bibitem{Bak05} H. Baker, Proc. Lond. Math. Soc. {\bf 3}, 24 (1905).
\bibitem{Hau06} F. Hausdorff, Ber. Verh. Saechs. Akad. Wiss. Leipzig {\bf 58}, 19 (1906).
\bibitem{Dyn47} E. B. Dynkin, Doklady Akademii Nauk SSSR {\bf 57}, 323 (1947).
\bibitem{AchBon12} R. Achilles, A. Bonfiglioli, Arch. Hist. Exact. Sci. {\bf 66}, 295 (2012). 
\bibitem{Dio18}  L. Di\'osi, J. Phys. A {\bf 51}, 365201 (2018).
\bibitem{Its07} M. Itskov, {\it Tensor Algebra and Tensor Analysis for Engineers}, Springer Verlag (2007).
\bibitem{Kel22} P. A. Kelly, {\it Mechanics Lecture Notes Part III: Foundations of Continuum Mechanics}, University of Auckland (2022). Online book available from http://homepages.engineering.auckland.ac.nz/~pkel015/\\SolidMechanicsBooks/index.html 
\bibitem{Spi71} M. Spivak, {\it Calculus On Manifolds}, Taylor\&Francis (1971).
\bibitem{supp} See Supplementary Material.
\bibitem{Cam98} J. E. Campbell, Proc. Lond. Math. Soc. {\bf29}, 14 (1898).
\bibitem{Poi00} H. Poincar\'e, Trans. Camb. Phil. Soc. {\bf18}, 220 (1900).
\bibitem{BrePet02} H.P. Breuer and F. Petruccione, {\it Theory of open quantum systems} (Oxford, Oxford University Press, 2002).
\bibitem{Fer17} L. Ferialdi Phys. Rev. A {\bf 95}, 020101(R) (2017); {\it ibid.} {\bf 95}, 069908(E) (2017).
\bibitem{Que14} N. Quesada and J. E. Sipe, Phys Rev. A {\bf 90}, 063840 (2014).
\bibitem{Que15} N. Quesada and J. E. Sipe, Phys Rev. Lett. {\bf 114}, 093903 (2015).

%\bibitem{DioFer14} L. Di\'osi, L. Ferialdi, Phys. Rev. Lett. {\bf 113}, 200403 (2014).
%\bibitem{Fer16} L. Ferialdi, Phys. Rev. Lett. {\bf 116}, 120402 (2016).
%\bibitem{Dio18a} L. Diosi, J. Russ. Las. Res. {\bf39}, 349 (2018).

\end{thebibliography}
\end{document}